\documentclass[12pt,a4paper]{article}
\usepackage{epsfig}
\usepackage{subfigure}
\usepackage{graphicx}
\usepackage{amsmath}
\usepackage{amssymb}
\usepackage{amsfonts}
\usepackage{amsmath}
\usepackage{cite}
\usepackage{multirow}

\begin{document}

\begin{titlepage}
\begin{flushright}
CERN-TH-2017-075
\end{flushright}
\begin{center}

{\large \bf {Leptonic Invariants, Neutrino Mass-Ordering and \\
the Octant of $\theta_{23}$}}

\vskip 1cm

G. C. Branco  $^{b, c}$ \footnote{gbranco@tecnico.ulisboa.pt}, 
M. N. Rebelo $^{b, c}$ \footnote{rebelo@tecnico.ulisboa.pt},
and J. I. Silva-Marcos $^b$\footnote{juca@cftp.ist.utl.pt},

\vspace{1.0cm}

{\it $^b$ Departamento de F\'\i sica and Centro de F\' \i sica Te\' orica
de Part\' \i culas (CFTP),
Instituto Superior T\' ecnico (IST), U. de Lisboa (UL), Av. Rovisco Pais, P-1049-001 
Lisboa, Portugal. \\
\it$^c$ Theory Department, CERN, CH 1211 Geneva 23, Switzerland}

\end{center}

\vskip 3cm

\begin{abstract}
We point out that leptonic weak-basis invariants are an important tool for the study 
of the properties of lepton flavour models. In particular, we show that appropriately 
chosen invariants can give a clear indication of whether a particular lepton flavour model
favours normal or inverted hierarchy for neutrino masses and what is the octant of 
$\theta_{23}$. These invariants can be evaluated in any conveniently chosen weak-basis 
and can also be expressed in terms of neutrino masses, charged lepton masses, 
mixing angles and CP violation phases. 
\end{abstract}

\end{titlepage}

\newpage

\section{Introduction}
Neutrino Physics is a very active field of research in Particle Physics, 
with a well established longterm program for future experiments. Although 
neutrino oscillations have provided solid evidence for leptonic mixing and 
for at least two non-vanishing neutrino masses, there are still some 
fundamental open questions. These include the establishment of the nature 
of neutrinos (Dirac or Majorana), determination of the pattern of neutrino 
masses (hierarchical or quasi-degenerate), settling of the ordering of 
neutrino masses (normal or inverted) and discovering leptonic CP violation.

We have reached a precision era for the measurement of leptonic mixing 
parameters and for the measurement of the squared mass differences  
of the three light neutrinos. Still, it is not yet clear whether it will 
be possible, in the near future, to determine the mass of the lightest 
neutrino and thus the scale of neutrino masses is not yet determined. However, 
it is by now established, both in laboratory experiments and
via astrophysical bounds, that light neutrinos can at most have 
masses of the order of one eV. Several of these open questions have profound 
implications for Astrophysics and Cosmology.

Recently, the Daya Bay Reactor Neutrino Experiment \cite{An:2012eh}  
measured with certainty, for the first time, a nonzero value for the 
smallest leptonic mixing angle, $\theta_{13}$. At that time it was 
already known that the two other leptonic mixing angles were large.
The fact that none of the three leptonic mixing angles vanishes 
opens up the possibility of observing leptonic CP violation 
of Dirac type in neutrino oscillation experiments. At present, there is a some 
likelihood indication for a Dirac phase of -$\pi/2$.
Until recently all experimental results were in agreement with 
$\theta_{23}$ corresponding to maximal mixing.
However, there is a new measurement by NOvA \cite{Adamson:2017qqn} reporting 
that this value is excluded at 2.6 $\sigma$ CL. 

On the theoretical side there have been many attempts at understanding 
the pattern 
of leptonic masses and mixing, through the introduction of family symmetries 
at the Lagrangian level or as symmetries of the leptonic mass matrices. 
In a bottom up 
approach, one may try to guess the family symmetries chosen by nature, 
from the input 
from experiment. 
One of the difficulties in pursuing this approach stems from the fact that 
the leptonic mass matrices change under weak-basis (WB) transformations. 
So even if there is a flavour symmetry chosen by nature, in what WB would 
the symmetry be evident?

In this paper we point out that leptonic WB invariants can be a very 
useful tool in the study of the pattern of leptonic masses and mixing, 
including leptonic 
CP violation. The paper is organised as follows. In the next section, we review 
leptonic CP-even and CP-odd WB invariants. In the CP-odd invariants, 
we include those which are sensitive to Dirac and Majorana-type CP violation. 
In the third section, we show how WB invariants provide a simple way of 
determining whether a given model favours
normal or inverted neutrino mass ordering and also what it predicts for 
the $\theta_{23}$ octant.
In section 4, we illustrate the usefulness of the WB invariants, by 
applying them to specific Ans\" atze proposed in the literature. 
The summary and conclusions are presented in the last section.

\section{Invariants and the Pattern of Leptonic Mixing and CP Violation}

\subsection{Introductory Remarks}

In the SM, the flavour structure of Yukawa couplings, in both the lepton and
quark sectors, is not constrained by gauge symmetry. As a result, fermion
masses and mixing are arbitrary. One may adopt a bottom up approach and
attempt at extracting from experiment some hint of a flavour symmetry. One
of the difficulties one encounters in this approach stems from the fact that
one has the freedom to make weak basis (WB) transformations under which the
flavour structure of Yukawa couplings change, but their physical content
remains invariant. Let us consider the SM and assume that lepton number is
violated by some physics beyond the SM, leading at low energies to an
effective Majorana neutrino mass matrix. The leptonic mass terms are: 
\begin{equation}
{\mathcal{L}}_{\mbox{mass}} = -\frac{1}{2} {\nu_L^{0}}^T C^{-1} m_\nu
\nu_L^{0} - \overline{\ell_L^0 } m_\ell \ell_R^0 + \text{h.c.}\ ,
\end{equation}
and the charged currents are: 
\begin{equation}
{\mathcal{L}}_W = - \frac{g}{\sqrt{2}} W_\mu^+ \overline{\ell_L^0 }
\gamma^\mu \nu_L^{0} + \text{h.c.}
\end{equation}
The WB transformations involving the leptonic fields are of the form: 
\begin{equation}
\nu_L^{0} \rightarrow V \nu_L^{0}, \quad \ell_L^0 \rightarrow V \ell_L^0,
\quad \ell_R^0 \rightarrow W \ell_R^0
\end{equation}
with $V$ and $W$ unitary $3 \times 3$ matrices. Under these transformations
the leptonic mass terms transform as: 
\begin{equation}
m_\nu \rightarrow V^T m_\nu V, \qquad m_\ell \rightarrow V^\dagger m_\ell W
\end{equation}
Leptonic mixing and CP violation in the leptonic sector are parametrised by
the Pontecorvo - Maki - Nakagawa - Sakata (PMNS) matrix, $U_{PMNS}$, which
contains three mixing angles and three CP violating phases, two of the
phases reflecting the Majorana character of neutrinos. \newline

Following the standard parametrisation \cite{Olive:2016xmw}  the matrix 
$U_{PMNS}$ can be denoted as:
\begin{eqnarray}
U_{PMNS}=\left(
\begin{array}{ccc}
c_{12} c_{13} & s_{12} c_{13} & s_{13} e^{-i \delta}  \\
-s_{12} c_{23} - c_{12} s_{23} s_{13}   e^{i \delta}
& \quad c_{12} c_{23}  - s_{12} s_{23}  s_{13} e^{i \delta} \quad 
& s_{23} c_{13}  \\
s_{12} s_{23} - c_{12} c_{23} s_{13} e^{i \delta}
& -c_{12} s_{23} - s_{12} c_{23} s_{13} e^{i \delta}
& c_{23} c_{13} 
\end{array}\right) \cdot  P
\label{std}
\end{eqnarray}
with P given by
\begin{equation}
P=\mathrm{diag} \ (1,e^{i\alpha_{21}}, e^{i\alpha_{31}})
\end{equation}
where $c_{ij} = \cos \theta_{ij}$, $s_{ij} = \sin \theta_{ij}$ the
angles $\theta_{ij} = [0, \pi/2]$, $\delta = [ 0. 2 \pi ] $ is a Dirac-type 
CP violating phase and $\alpha_{21}$, $\alpha_{31}$ denote phases associated 
to the Majorana character of neutrinos.  Neutrino oscillation experiments
are sensitive to the  mixing parameters with the exception of the 
CP violating phase $\alpha_{21}$, $\alpha_{31}$. There is no loss of generality
in adopting the convention that $\theta_{ij}$ are all in the first quadrant.

In Table 1 we summarise the present knowledge concerning neutrino
masses and leptonic mixing. In the literature, there are three global
phenomenological fits on $\theta_{12}$, $\theta_{23}$, $\theta_{13}$,  
and $\delta$ \cite{Forero:2014bxa}, \cite{Capozzi:2013csa}, 
\cite{Gonzalez-Garcia:2015qrr}. 
The specific bounds vary slightly from reference to reference.
For definiteness we present those of Ref. \cite{Forero:2014bxa}.

\begin{center}
\begin{table}[h]
\caption{Neutrino oscillation parameter summary, taken from 
Ref.~\protect\cite{Forero:2014bxa}. 
For $\Delta m^2_{31}$, 
$\sin^2 \protect\theta_{23}$, $\sin^2 \protect\theta_{13}$, and $\protect
\delta$ the upper (lower) row corresponds to normal (inverted) neutrino mass
hierarchy. ${}^a$There is a local minimum in the first octant, $\sin^2 
\protect\theta_{23}= 0.467 $ with $\Delta \protect\chi^2 =0.28$ with respect
to the global minimum.}
\label{reps}
\begin{center}
\begin{tabular}{ccc}
\hline\hline
Parameter & Best fit & $1 \sigma $ range \\ \hline
$\Delta m^2_{21}$ $[10^{-5} eV^2 ] $ & 7.60 & 7.42 -- 7.79 \\ 
$|\Delta m^2_{31}|$ $[10^{-3} eV^2 ] (NH)$ & 2.48 & 2.41 -- 2.53 \\ 
$|\Delta m^2_{31}|$ $[10^{-3} eV^2 ] (IH) $ & 2.38 & 2.32 -- 2.43 \\ 
$\sin^2 \theta_{12}$ & 0.323 & 0.307 -- 0.339 \\ 
$\sin^2 \theta_{23} (NH)$ & 0.567 & $0.439^a$ -- 0.599 \\ 
$\sin^2 \theta_{23} (IH)$ & 0.573 & 0.530 -- 0.598 \\ 
$\sin^2 \theta_{13}$ (NH) & 0.0234 & 0.0214 --0.0254 \\ 
$\sin^2 \theta_{13}$ (IH) & 0.0240 & 0.0221 -- 0.0259 \\ 
$\delta$ (NH) & 1.34 $\pi $ & 0.96 --1.98 $\pi$ \\ 
$\delta$ (IH) & 1.48 $\pi$ & 1.16 --1.82 $\pi$ \\ \hline
\end{tabular}
\end{center}
\end{table}
\end{center}

Neutrino oscillations give information about differences of squared masses:
\begin{equation}
\Delta m^2_{21} \equiv \Delta^2_{21} \equiv m^2_2 - m^2_1 , \quad
\Delta m^2_{31} \equiv \Delta^2_{31} \equiv m^3_2 - m^2_1 
\end{equation}
The sign of $\Delta m^2_{31}$ is not yet known. The best fit values of some of the
parameters listed in Table 1 depend on the sign of $\Delta m^2_{31}$. For a
positive sign the ordering is called normal (NH), for a negative sign the 
ordering is called inverted (IH). The association of the terms normal and inverted 
to each one of the signs reflects a prejudice, since 
from a theoretical point of view, no ordering can a priori be 
considered more natural, as discussed in Ref.~\cite{Branco:2011aa}.

At this stage, it is worth recalling the main differences between rephasing
invariant quantities in the cases of Majorana and Dirac neutrinos.
Let us start by considering unitarity triangles, assuming that the 
$U_{PMNS}$ is a 3x3 unitary matrix. It is well known that there are many
frameworks, including for example the seesaw type one \cite{Minkowski:1977sc},
\cite{Yanagida:1979as}, \cite{Levy:1980ws}, \cite{GellMann:1980vs}, \cite{Mohapatra:1979ia}
where this is not
exactly true, since there are small deviations from 3x3 unitarity. With a
unitary $U_{PMNS}$ , one has six leptonic unitarity triangles, three
corresponding to orthogonality of rows and another three for orthogonality
of columns. The triangles corresponding to orthogonality of rows are often
called Dirac triangles and are very similar to the unitarity triangles in
the quark sector. Under rephasing of the charged lepton fields, the leptonic
Dirac triangles rotate and thus the direction of their sides have no
physical meaning. Analytically, they correspond to quantities like $%
\arg(U_{e1} U^*_{\mu1})$ which are not rephasing invariant. The phases which
are physically meaningful in these Dirac triangles are the internal angles
of the triangles which analytically correspond to the arguments of invariant
leptonic quartets like $(U_{e2} U_{\mu3} U^*_{e3}U^*_{\mu2})$. In the
Majorana triangles, one encounters a very different situation \cite%
{AguilarSaavedra:2000vr}. In these triangles the directions of the sides are
physically meaningful and do not change under the rephasing of the charged
lepton fields. Recall that one cannot rephase Majorana neutrinos.
Analytically these directions correspond to
rephasing-invariant biliniars like $(U_{e1} U^*_{e2})$ . Therefore, the most
rigorous definition of Majorana phases is that they correspond to arguments
of the rephasing invariant bilinears $(U_{\ell j} U^*_{\ell k})$. 
It can be seen that, independently of
unitarity, there are only six independent Majorana phases in
a $3 \times 3$ $U_{PMNS}$. Assuming
unitarity, it has been shown that from the knowledge of six independent
Majorana phases one can construct the full PMNS matrix, including moduli and
phases \cite{Branco:2008ai}.

\subsection{Leptonic Weak-Basis Invariants}
In this subsection, we describe the WB invariants which can fix the
lepton mixing and CP violation in the leptonic sector. We consider 
WB invariants written in terms of the charged lepton mass matrix and the 
effective neutrino mass matrix and not WB invariants written in terms
of the neutrino mass matrices \cite{Pilaftsis:1997jf}, \cite{Branco:2001pq}, 
\cite{Davidson:2003yk}, \cite{Branco:2005jr}, \cite{Wang:2014lla}
appearing in the framework of the seesaw mechanism.

It can be shown that the following four weak basis
(WB) invariants completely define four independent moduli of 
$U_{PMNS}$ \cite{Branco:1987mj}: 
\begin{eqnarray}
I_1 = \mbox{Tr}\lbrack H_{\ell}\ H_{\nu}], \qquad I_2 = \mbox{Tr}\lbrack
H_{\ell}^2\ H_{\nu}],  \label{15}\\
I_3 = \mbox{Tr}\lbrack H_{\ell}\ H_{\nu}^2], \qquad I_4 = \mbox{Tr}\lbrack
H_{\ell}^2\ H_{\nu}^2] \label{16}
\end{eqnarray}
where $H_{\nu} = m_\nu^* \ m_\nu^T$ and $H_{\ell} = m_\ell \ m_\ell^\dagger$.
These four WB invariants are physical quantities and can be expressed in
terms of charged lepton and neutrino masses and moduli of $U_{PMNS}$. From
the knowledge of the four invariants and the charged lepton and neutrino
masses, one can derive all the moduli of $U_{PMNS}$, using 3x3 unitarity.
From the knowledge of the moduli one can then readily evaluate the common
area of all unitarity triangles which in turn gives the strength of leptonic
CP violation of the Dirac type. This is entirely analogous to the situation
in the quark sector \cite{Branco:1987mj}. 
Although the four invariants of Eqs~(\ref{15}) and (\ref{16}), together with 
$3 \times 3$ unitarity completely fix the moduli of $U_{PMNS}$ and
the strength of leptonic CP violation of Dirac-type, there is still a
two-fold ambiguity, since the sign of CP violation is not fixed. 
This ambiguity can be lifted by calculating \cite{Bernabeu:1986fc}:
\begin{equation}
I^{CP} \equiv \mbox{Tr}\  [m_{\nu}^* \cdot  m_{\nu}^T, \ h_\ell]^3 
\end{equation}
At this stage it should be pointed out that there is an important
difference between the lepton and quark sectors. While in the quark sector
one can overdetermine the CKM matrix from experiment, in the case of the
lepton sector with Majorana neutrinos, one cannot completely determine $%
U_{PMNS}$ from feasible experiments. This is related to the appalling fact,
emphasised by Glashow et al, \cite{Frampton:2002yf} that it is not possible
to fully reconstruct the neutrino mass matrix from feasible experiments. It
is instructive to write explicitly the strength of Dirac type CP violation
in terms of four independent moduli of $U_{PMNS}$. Choosing as independent
moduli $U_{e2}$, $U_{e3}$, $U_{\mu 3}$, $U_{\mu 2}$, one obtains 
\cite{Branco:1987mj}: 
\begin{eqnarray}
\mbox{Im Q} \equiv \mbox{Im}\left( U_{e2}\ U_{\mu 3}\ U_{e3}^* \ U_{\mu 2}^*
\right) = \sqrt{|U_{e2}|^2 \ |U_{\mu 3}|^2 \ |U_{e3}|^2 \ |U_{\mu 2}|^2 - R^2%
}, \\
R = \left( 1- |U_{e2}|^2 - |U_{\mu 3}|^2 - |U_{e3}|^2 - |U_{\mu 2}|^2 +
|U_{e2}|^2 \ |U_{\mu 3}|^2 + |U_{e3}|^2 \ |U_{\mu 2}|^2 \right)/ 2
\end{eqnarray}
Experimentally one can extract information on the real and imaginary 
parts of such quartets from neutrino oscillation experiments.
This is to be compared to the
quark sector where one can choose as input moduli $|V_{us}|$, $|V_{ub}|$, 
$|V_{cb|} $ and $|V_{td}|$. The first three moduli can be measured from strange
particles, and B-meson decays, while $|V_{td}|$ can be measured from $B_d -- 
\overline{B_d}$ mixing. Of course, the measurement of $|V_{td}|$ through this
meson mixing can be affected by New Physics contributions to
this mixing. In spite of the scarcity of leptonic measurements, 
a nice aspect of the leptonic sector
is the absence of ``hadronic uncertainties". For example, it is remarkable
that the present experimental measurement $|U_{e3}|$ has a smaller percent
error than the measurement of $|V_{ub}|$, in spite of the enormous effort from
both theorists and experimentalists to measure $|V_{ub}|$.

So far, we have only considered WB invariants which fix the strength of CP
violation of Dirac-type. In the case of Majorana neutrinos one has two extra
phases which, as emphasised before, have to do with the fact that 
for Majorana neutrinos there is only freedom to rephase the charged 
leptons and therefore the phase of the bilinear  $(U_{\ell j} U^*_{\ell k})$
has physical meaning.

It is possible to derive WB invariant CP-odd conditions sensitive to the
three CP violating phases present in Eq.~(\ref{std}). This was done in 
Ref.~\cite{Branco:1986gr}
where it was shown that the following set of conditions
is necessary and sufficient for CP invariance
in the case of three generations, for nonzero
and nondegenerate masses:
\begin{eqnarray}
\mbox{Im}\  \mbox{Tr} \ [h_{\ell} \cdot m_{\nu}^*\cdot m_{\nu} \cdot m_{\nu}^* \cdot h_{\ell}^* 
\cdot m_{\nu} ]  
\  = 0  \label{asd} \\
\mbox{Im}\  \mbox{Tr} \ [ h_{\ell} \cdot( m_{\nu}^*\cdot m_{\nu})^2 \cdot (m_{\nu}^* 
\cdot h_{\ell}^* \cdot m_{\nu}) ] = 0\\
\mbox{Im}\  \mbox{Tr} [ h_{\ell} \cdot( m_{\nu}^*\cdot m_{\nu})^2 \cdot (m_{\nu}^* 
\cdot h_{\ell}^* \cdot m_{\nu})(m_{\nu}^* \cdot m_{\nu}] = 0\\
\mbox{Im}\  \mbox{Tr} [(m_{\nu} \cdot h_{\ell} \cdot m_{\nu}^*) + (h_{\ell}^* \cdot m_{\nu}
\cdot m_{\nu}^*)] = 0
\end{eqnarray}
Selecting a minimal set of necessary and sufficient conditions for CP invariance
is not trivial and was provided later on, in Ref.~ \cite{Dreiner:2007yz}:
\begin{eqnarray}
\mbox{Tr}\  [m_{\nu}^* \cdot  m_{\nu}^T, \ h_\ell]^3 = 0    \\
\mbox{Tr}\  [m_{\nu} \cdot h_\ell \cdot  m_{\nu}^*, h_\ell^*]^3 = 0 \\
\mbox{Im} \mbox{Tr} \ (h_{\ell} \cdot m_{\nu}^*\cdot m_{\nu} \cdot m_{\nu}^* 
\cdot h_{\ell}^* \cdot m_{\nu}) = 0 
\end{eqnarray}
The first of these three equations is similar to the condition derived
for the quark sector. It is sensitive to the Dirac-type phase and insensitive
to the Majorana-type phases. The second and third equations are
sensitive to both Dirac and Majorana type phases. The second equation was 
first derived in Ref~\cite{Branco:1998bw} in the context of three degenerate 
neutrinos which, as was shown, still allows for leptonic mixing and Majorana
type CP violation. The third equation coincides with Eq.~(\ref{asd}) 
which was derived for the first time in Ref.~\cite{Branco:1986gr} where 
it was shown that it is the necessary and sufficient condition for
CP conservation in the case of two generations. Recall that for 
two generations only, Majorana-type CP violation can occur.

\section{Invariants sensitive to neutrino mass ordering and the $\protect%
\theta_{23}$ octant}

In the previous section we summarised important information on weak basis
invariants that was already known and that have proved to be extremely
useful. In this section we discuss a set of new invariants sensitive to the
neutrino mass ordering and to the octant in which the angle $\theta_{23}$
lies. One important feature of the new invariants is the fact that their
building blocks are analogous to the invariants found in Ref.\cite{Branco:1987mj}
for the quark sector.

\subsection{The neutrino mass ordering}

One of the outstanding questions is the neutrino mass ordering. It is not
yet known whether or not the mass of $m_3$ is higher than the mass of $m_1$
(and of $m_2$). This refers to the neutrino mass ordering associated to the 
$U_{PMNS}$ angles as given in Table 1. The scale of neutrino masses is also
not yet fixed. The highest hierarchy is obtained when the lightest neutrino
mass is close to zero. However, it may happen that the three neutrino masses
are almost degenerate. Almost degeneracy requires the mass scale to be
higher than the square root of $|\Delta m^2_{31}|$. \newline

Depending on the neutrino mass ordering, it is useful to consider different
parametrisations for the neutrino masses. For normal ordering the following
parametrisation is useful: 
\begin{equation}
\begin{array}{l}
m_{1}^{2}=\Delta ^{2}\ x \\ 
\\ 
m_{2}^{2}=\Delta ^{2}\ \left( x+\epsilon \right)  \\ 
\\ 
m_{3}^{2}=\Delta ^{2}\ \left( x+1\right) 
\end{array}%
\qquad ;\qquad 
\begin{array}{c}
\Delta ^{2}=|\Delta _{31}^{2}| \\ 
\\ 
\epsilon =\frac{\Delta _{21}^{2}}{|\Delta _{31}^{2}|}
\end{array}%
\qquad ;\qquad 0\leq x  \label{no}
\end{equation}%
For inverted ordering, we use the following parametrisation for the neutrino
masses: 
\begin{equation}
\begin{array}{l}
m_{1}^{2}=\Delta ^{2}\ \left( x^{\prime }+1\right)  \\ 
\\ 
m_{2}^{2}=\Delta ^{2}\ \left( x^{\prime }+1+\epsilon \right)  \\ 
\\ 
m_{3}^{2}=\Delta ^{2}\ x^{\prime }%
\end{array}%
\qquad ;\qquad 0\leq x^{\prime }  \label{iv}
\end{equation}%
In can be easily shown that the sign of the following WB invariant: 
\begin{equation}
\tilde{I_{1}}\equiv Tr[H_{\ell }\ H_{\nu }]-\frac{1}{3}Tr[H_{\ell
}]Tr[H_{\nu }]
\end{equation}
indicates the ordering of the neutrino masses. Since $\tilde{I_1}$ is a WB
invariant it can be computed in any particular WB. It is instructive to
compute $\tilde{I_{1}}$ in the basis where the charged lepton mass matrix is
diagonal. In this basis we have: 
\begin{equation}
H_{\ell }=\mathrm{diag}\ (m_{e}^{2},\ m_{\mu }^{2},\ m_{\tau }^{2}),\qquad
H_{\nu }=U_{PMNS}\ d_{\nu }^{2}\ U_{PMNS}^{\dagger }
\end{equation}
with $d_{\nu }^{2} = \mathrm{diag} \ (m_{1}^{2},\  m_{2}^{2},\  m_{3}^{2})$.
This allows us to express $\tilde{I_{1}}$ in terms of physical observables:
\begin{eqnarray}
Tr[H_{\nu }] &=&m_{1}^{2}+m_{2}^{2}+m_{3}^{2},\qquad Tr[H_{\ell
}]=m_{e}^{2}+m_{\mu }^{2}+m_{\tau }^{2} \\
Tr[H_{\ell }\ H_{\nu }] &=&m_{e}^{2}m_{k}^{2}\left\vert U_{1k}\right\vert
^{2}+m_{\mu }^{2}m_{k}^{2}\left\vert U_{2k}\right\vert ^{2}+m_{\tau
}^{2}m_{k}^{2}\left\vert U_{3k}\right\vert ^{2}
\end{eqnarray}%
where summation in $k$ is implied and the $U_{ij}$ stand for the entries of $%
U_{PMNS}$. Then, it is straightforward to calculate $\tilde{I_{1}}$. We find:
\begin{equation}
\begin{array}{r}
\tilde{I_{1}}=m_{\tau }^{2}\left[ \Delta _{31}^{2}\left( \left\vert
U_{33}\right\vert ^{2}-\frac{1}{3}\right) +\Delta _{21}^{2}\left( \left\vert
U_{32}\right\vert ^{2}-\frac{1}{3}\right) \right] + \\ 
\\ 
+m_{\mu }^{2}\left[ \Delta _{31}^{2}\left( \left\vert U_{23}\right\vert ^{2}-%
\frac{1}{3}\right) +\Delta _{21}^{2}\left( \left\vert U_{22}\right\vert ^{2}-%
\frac{1}{3}\right) \right] + \\ 
\\ 
+m_{e}^{2}\left[ \Delta _{31}^{2}\left( \left\vert U_{13}\right\vert ^{2}-%
\frac{1}{3}\right) +\Delta _{21}^{2}\left( \left\vert U_{12}\right\vert ^{2}-%
\frac{1}{3}\right) \right] 
\end{array}
\label{i1}
\end{equation}
Taking into account the value of charged lepton masses:
\begin{equation}
m_{e}^{2}=2.6\times 10^{-7}\mathrm{GeV}^{2},\quad m_{\mu }^{2}=1.1\times
10^{-2}\mathrm{GeV}^{2}\quad m_{\tau }^{2}=3\mathrm{GeV}^{2},
\end{equation}
it is clear that we can safely neglect the terms in $m_{e}^{2}$ and $m_{\mu }^{2}$ 
in the determination of the sign of $\tilde{I_{1}}$. 
Furthermore, experimentally we
know that $\left\vert U_{33}\right\vert ^{2} > 1/3$. Therefore, it follows that
the sign of $\tilde{I_{1}}$ gives the sign of $\Delta _{31}^{2}$. It is
interesting to note that for  exact tribimaximal mixing \cite{Harrison:2002er}
which leads to leptonic mixings close to the experimental values: 
\begin{equation}
U_{TBM}=\left( 
\begin{array}{ccc}
\frac{2}{\sqrt{6}} & \frac{1}{\sqrt{3}} & 0 \\ 
\frac{1}{\sqrt{6}} & -\frac{1}{\sqrt{3}} & \frac{1}{\sqrt{2}} \\ 
\frac{1}{\sqrt{6}} & -\frac{1}{\sqrt{3}} & -\frac{1}{\sqrt{2}}%
\end{array}
\right)
\end{equation}
one has the following expression for $\tilde{I_{1}}$: 
\begin{equation}
\tilde{I_{1}} \equiv Tr[H_{\ell }\ H_{\nu }]-\frac{1}{3}Tr[H_{\ell }]Tr[H_{\nu }]
\simeq \frac{1}{6}
m_{\tau }^{2}\ \Delta _{31}^{2}
\end{equation}

\subsection{The octant of $\protect\theta_{23}$}

Despite great experimental progress in the determination of the neutrino
oscillation parameters, two of these still remain poorly known - the
atmospheric mixing angle $\theta _{23}$ and the CP violating Dirac type
phase $\delta $. The forthcoming neutrino oscillation experiments are
expected to significantly improve their measurements. Concerning the angle $%
\theta _{23}$ there are two degenerate solutions known as the octant problem 
\cite{Fogli:1996pv}. One, the lower octant solution corresponds to $\theta
_{23} < \pi /4$, the other, the higher octant solution
corresponds to $\theta _{23} > \pi /4$. The recent measurement
of the angle $\theta _{13}$ \cite{An:2012eh} and the fact that it is not too
small gives grounds for optimism concerning the possibility of resolving the
octant issue in forthcoming neutrino experiments \cite{Minakata:2002jv}, 
\cite{Chatterjee:2017irl}.

It is remarkable that there is a WB invariant which is sensitive 
to the $\theta _{23}$ octant, namely: 
\begin{equation}
\tilde{I_{2}}\equiv Tr[H_{\ell }]\ Tr[H_{\ell }^{2}\ H_{\nu }]-Tr[H_{\ell
}^{2}]Tr[H_{\ell }\ H_{\nu }]  \label{i22}
\end{equation}%
We find
\begin{eqnarray}
\tilde{I_{2}}=\Delta _{31}^{2}  [ 
m_{\tau }^{2}m_{\mu }^{2}\left( m_{\tau }^{2}-m_{\mu }^{2}\right) \left(
\left\vert U_{33}\right\vert ^{2}-\left\vert U_{23}\right\vert ^{2}\right)
+m_{\tau }^{2}m_{e}^{2}\left( m_{\tau }^{2}-m_{e}^{2}\right) \left(
\left\vert U_{33}\right\vert ^{2}-\left\vert U_{13}\right\vert ^{2}\right) +
\nonumber \\ 
 +  m_{\mu }^{2}m_{e}^{2}\left( m_{\mu }^{2}-m_{e}^{2}\right) \left( \left\vert
U_{23}\right\vert ^{2}-\left\vert U_{13}\right\vert ^{2}\right) ] + \nonumber \\ 
\nonumber \\ 
+\Delta _{21}^{2} [ 
m_{\tau }^{2}m_{\mu }^{2}\left( m_{\tau }^{2}-m_{\mu }^{2}\right) \left(
\left\vert U_{32}\right\vert ^{2}-\left\vert U_{22}\right\vert ^{2}\right)
+m_{\tau }^{2}m_{e}^{2}\left( m_{\tau }^{2}-m_{e}^{2}\right) \left(
\left\vert U_{32}\right\vert ^{2}-\left\vert U_{12}\right\vert ^{2}\right) 
\nonumber\\ 
+m_{\mu }^{2}m_{e}^{2}\left( m_{\mu }^{2}-m_{e}^{2}\right) \left( \left\vert
U_{22}\right\vert ^{2}-\left\vert U_{12}\right\vert ^{2}\right) ] 
\label{ii2}
\end{eqnarray}
It is clear that the sign of $\tilde{I_{2}}$ gives the sign of $\left(
\left\vert U_{33}\right\vert ^{2}-\left\vert U_{23}\right\vert ^{2}\right)$,
once the sign of $\Delta _{31}^{2}$ is known from $\tilde{I_{1}}$. In Table 2,
we illustrate how the knowledge of the sign of $\tilde{I_{1}}$ and $\tilde{I_{2}}$
determines the neutrino mass ordering as well as the $\theta_{23}$ octant.

\begin{table}[htb]
\caption{Combination of the two invariants. NO stands for normal 
ordering, IO for inverted ordering}
\label{Table:combination}
\begin{center}
\begin{tabular}{|c|c|c|}
\cline{2-3}
\multicolumn{1}{c|}{} &  &  \\
\multicolumn{1}{c|}{} & $\tilde{I_{2}} > 0$ & $\tilde{I_{2}} < 0$ \\
\hline
  & &   \\
$\tilde{I_{1}} > 0$ & NO, $\theta_{23} < \pi/4$ & 
NO, $\theta_{23} > \pi/4$\\
\hline
  & &   \\
$\tilde{I_{1}} < 0$ & IO, $\theta_{23} > \pi/4$ & 
IO, $\theta_{23} < \pi/4$\\ 
\hline 
\hline
\end{tabular}
\end{center}
\end{table}

\section{Application to specific Ans\"{a}tze for Leptonic masses}
Recently, various analysis of the prediction of neutrino mass textures 
have been presented in  the literature. Typically a random scan 
\cite{Cebola:2015dwa}, \cite{Cebola:2016jhz} is performed,
with the input of the parameters of the Ans\" atze, leading to the 
determination of the various predictions of the  Ans\" atze for
a selected number of physical parameters. The invariants $\tilde{I_{1}}$
and $\tilde{I_{2}}$ are a complementary tool for these analysis,
providing a simple determination of the favoured neutrino mass
ordering and the octant of $\theta_{23}$.

For illustrative purposes, we use the two invariants $\tilde{I_i}$ in 
the case of two specific Ans\" atze, studied in 
Ref.~\cite{Frampton:2002yf}, which predict
a definite neutrino mass ordering.

In Ref.~\cite{Frampton:2002yf} the authors considered neutrino mass matrices 
with the maximal allowed number of zero textures in the WB where the charged 
lepton mass matrix is already diagonal. They concluded that no more than 
two independent zero textures were viable. Furthermore, out of the fifteen
different choices only seven could accommodate the known experimental 
constraints. Texture zeros in $m_\nu$ lead to predictions. In both examples 
the neutrino mass ordering is fixed by the chosen texture and therefore, 
as we are going to show the sign of $\tilde{I_1}$ is fixed. \\

In the case of texture $A_1$ defined as \cite{Frampton:2002yf}:
\begin{equation}
m_{\nu }=\left( 
\begin{array}{ccc}
0 & 0 & a \\ 
0 & b & c \\ 
a & c & d
\end{array}
\right)   \label{a1}
\end{equation}
the computation of the invariants  $\tilde{I_1}$ and  $\tilde{I_2}$ leads
in leading order to:
\begin{small}
\begin{eqnarray} 
\tilde{I_{1}}& \simeq &\frac{1}{3}m_{\tau }^{2}\left( \left\vert a\right\vert
^{2}+\left\vert c\right\vert ^{2}+2\left\vert d\right\vert ^{2}-\left\vert
b\right\vert ^{2}\right) \simeq \frac{1}{6} m_{\tau }^{2} \Delta_{31}^2 
\label{prim}\\  
\tilde{I_{2}}& \simeq & m_{\tau }^{2}m_{\mu }^{2}\left( m_{\tau
}^{2}-m_{\mu }^{2}\right) \left( \left\vert a\right\vert ^{2}+\left\vert
d\right\vert ^{2}-\left\vert b\right\vert ^{2}\right) \simeq
\Delta _{31}^{2}  
m_{\tau }^{2}m_{\mu }^{2}\left( m_{\tau }^{2}-m_{\mu }^{2}\right) \left(
\left\vert U_{33}\right\vert ^{2}-\left\vert U_{23}\right\vert ^{2}\right)
\label{seg}
\end{eqnarray}
\end{small}
From Eq.~(\ref{seg}) we get to a good approximation that:
\begin{equation}
\left\vert b\right\vert ^{2} = \left\vert a\right\vert ^{2}+\left\vert
d\right\vert ^{2} - \left(
\left\vert U_{33}\right\vert ^{2}-\left\vert U_{23}\right\vert ^{2}\right) 
\ \Delta _{31}^{2} 
\end{equation}
Replacing $\left\vert b\right\vert ^{2}$ into Eq.~(\ref{prim}) we get:
\begin{equation}
\left\vert c\right\vert ^{2}+\left\vert d\right\vert ^{2} \simeq 
\left[ \frac{1}{2} + \left(
\left\vert U_{33}\right\vert ^{2}-\left\vert U_{23}\right\vert ^{2}\right) \right] 
\Delta _{31}^{2} 
\end{equation}
The lefthand side is positive definite and we know experimentally (see Table 1) that 
the term in brackets on the righthand side cannot be negative, so we conclude that 
in this case $\Delta _{31}^{2} $ must be positive. \\

Another interesting example is case C, which corresponds to the following 
texture \cite{Frampton:2002yf}:
\begin{equation}
m_{\nu }=\left( 
\begin{array}{ccc}
a & c_1 & c_2 \\ 
c_1 & 0 & c_3 \\ 
c_2 & c_3 & 0
\end{array}
\right)   \label{a1}
\end{equation}
Computing the invariants  $\tilde{I_1}$ and  $\tilde{I_2}$ we obtain 
for the leading order terms:
\begin{eqnarray} 
\tilde{I_{1}}& \simeq &\frac{1}{3}m_{\tau }^{2}\left( - \left\vert a\right\vert
^{2}- 2\left\vert c_1\right\vert ^{2} + \left\vert c_2\right\vert ^{2}
+ \left\vert c_3\right\vert ^{2} \right) \simeq \frac{1}{6} m_{\tau }^{2} \Delta_{31}^2 
\label{prim1}\\  
\tilde{I_{2}}& \simeq & m_{\tau }^{2}m_{\mu }^{2}\left( m_{\tau
}^{2}-m_{\mu }^{2}\right) \left( \left\vert c_2\right\vert ^{2}- \left\vert
c_1 \right\vert^{2}  \right)\simeq
\Delta _{31}^{2}  
m_{\tau }^{2}m_{\mu }^{2}\left( m_{\tau }^{2}-m_{\mu }^{2}\right) \left(
\left\vert U_{33}\right\vert ^{2}-\left\vert U_{23}\right\vert ^{2}\right)
\label{seg2}
\end{eqnarray} 
In this case it is not straightforward to apply the previous procedure since 
$\tilde{I_1}$ and $\tilde{I_2}$ cannot be simply related. However close to
tribimaximal mixing it can be shown that:
\begin{eqnarray}
\left\vert a\right\vert ^{2} \simeq \left\vert c_3\right\vert ^{2} \qquad
\left\vert c_1\right\vert ^{2} \simeq \left\vert c_2\right\vert ^{2}
\end{eqnarray}
so that:
\begin{eqnarray}
m_1^2 \simeq \left\vert c_3\right\vert ^{2} + 2 \left\vert c_1\right\vert ^{2} \quad
m_2^2 \simeq \left\vert c_3\right\vert ^{2} + 2 \left\vert c_1\right\vert ^{2} \quad
m_3^2 \simeq \left\vert c_3\right\vert ^{2} 
\end{eqnarray}
and 
\begin{equation}
\tilde{I_1} \simeq \frac{1}{3}m_{\tau }^{2}\left( - \left\vert c_1\right\vert ^{2} \right)
<0
\end{equation}
which indicates that this texture favours inverted order.

\section{Conclusions}
We have emphasised that WB invariants can play an important role
in the study of lepton masses and mixing, including CP violation. 
The great advantage of these invariants stems from the fact that they can be 
directly evaluated in any conveniently chosen weak basis, without involving the 
diagonalisation of complex mass matrices. The invariants are physical quantities 
and can be expressed in terms neutrino masses, charged lepton masses, 
mixing angles and CP violating phases. We first review the four WB invariants 
which, together with the assumption of $3 \times 3$ unitarity of 
$U_{PMNS}$ matrix, can completely fix all the moduli of $U_{PMNS}$. 
From these moduli, one can evaluate the common area of all leptonic 
unitarity triangles. This area gives the strength of leptonic CP 
violation of Dirac type, but it does not fix the sign of CP violation. 
This sign can be fixed by a CP-odd leptonic WB invariant.
We have also described the WB invariants which can probe CP violation of 
Majorana type, emphasising that this CP violation has to do with the 
orientation of  the sides of Majorana-type unitarity triangles. For Majorana 
neutrinos, this orientation is physically meaningful and is associated 
to the arguments of bilinears of $U_{PMNS}$ 
matrix elements. Finally, we have shown that one can construct additional 
WB invariants which
can determine whether the neutrino mass ordering is normal or inverted 
and also determine the octant of $\theta_{23}$. These invariants are then 
used to study specific
texture-zero Ans\" atze for the neutrino mass matrices.

\section*{Acknowledgments}

This work was partially supported by Funda\c c\~ ao para a Ci\^ encia e a
Tecnologia (FCT, Portugal) through the projects CERN/FIS-NUC/0010/2015,
CFTP-FCT Unit 777 \newline
(UID/FIS/00777/2013) which are partially funded through POCTI (FEDER),
COMPETE, QREN and EU.

\end{document}